\documentclass[9pt]{extarticle}

\usepackage{microtype}
\usepackage[utf8]{inputenc}
\usepackage{longtable}
\usepackage{eurosym}
\usepackage{url}
\usepackage{amssymb}
\usepackage{pifont}
\usepackage[square,numbers]{natbib}
\usepackage{fontawesome5}
\usepackage{colortbl}
\usepackage{multirow}
\usepackage{lipsum}
\usepackage{amsthm}
\usepackage{mathtools}
\usepackage{xfrac}
\usepackage[export]{adjustbox}
\usepackage{longtable}
\usepackage{makecell}
\usepackage{booktabs}
\usepackage[table,xcdraw]{xcolor}
\usepackage{amssymb}
\usepackage{fancyhdr}

\definecolor{cai_primary}{HTML}{4C9A99}
\definecolor{cai_secondary}{HTML}{307FE2}
\definecolor{cai_accent}{HTML}{1D8348}
\definecolor{cai_dark}{HTML}{3F4444}
\definecolor{cai_light}{HTML}{F5F5F5}
\definecolor{cai_purple}{HTML}{8A4FFF}
\definecolor{pending_gray}{HTML}{888888}

\newlength{\tablewidth}
\newcommand{\robotimage}{\includegraphics[width=\tablewidth]{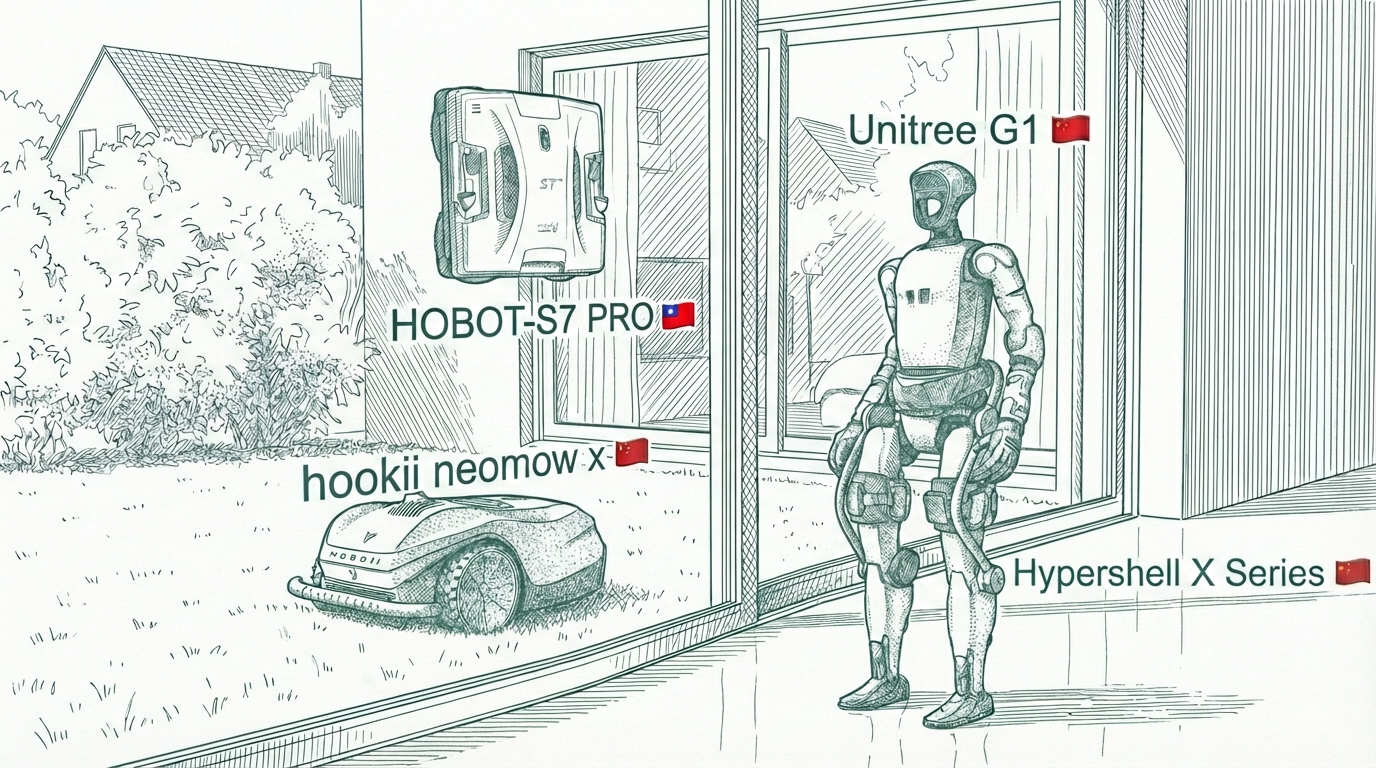}\vspace{0.8em}}

\usepackage{graphicx}
\usepackage{subcaption}
\usepackage{verbatim}
\usepackage{placeins}
\usepackage{mdframed}
\usepackage{hyperref}
\hypersetup{
    colorlinks=true,
    urlcolor=cai_secondary,
    linkcolor=cai_secondary,
    filecolor=cai_accent,
    citecolor=cai_secondary,
}
\usepackage{fancyvrb}
\usepackage{fixltx2e}
\usepackage{bera}

\usepackage{tikz}
\usetikzlibrary{arrows.meta,positioning,shapes.geometric,calc,patterns}
\usepackage{pgfplots}
\pgfplotsset{compat=1.16}
\pgfplotsset{%
  sharpbar/.style={%
    ybar,
    draw=black,
    line width=1.2pt,
    rounded corners=0.5pt,
    preaction={%
      transform canvas={xshift=4pt,yshift=-2pt},
      draw=none,
      fill=black!60,
      rounded corners=0.5pt
    }%
  },
  sharpbarh/.style={%
    xbar,
    draw=black,
    line width=1.2pt,
    rounded corners=0.5pt,
    preaction={%
      transform canvas={xshift=4pt,yshift=-2pt},
      draw=none,
      fill=black!60,
      rounded corners=0.5pt
    }%
  }%
}

\renewcommand{\headrulewidth}{0.4pt}
\renewcommand{\footrulewidth}{0.4pt}
\renewcommand{\headrule}{\hbox to\headwidth{\color{cai_primary}\leaders\hrule height \headrulewidth\hfill}}
\renewcommand{\footrule}{\hbox to\headwidth{\color{cai_dark}\leaders\hrule height \footrulewidth\hfill}}

\usepackage{algorithm}
\usepackage{listings}
\usepackage{geometry}

\usepackage{caption,setspace}
\usepackage{titlesec}
\titleformat{\section}
  {\normalfont\Large\bfseries\color{cai_primary}}
  {\thesection}
  {1em}
  {}
  [\titlerule]

\titleformat{\subsection}
  {\normalfont\large\bfseries\color{cai_dark}}
  {\thesubsection}
  {1em}
  {}

\titleformat{\subsubsection}
  {\normalfont\normalsize\bfseries\color{cai_dark}}
  {\thesubsubsection}
  {1em}
  {}

\definecolor{human_color}{HTML}{173C47}

\usepackage{authblk}

\renewcommand\Affilfont{\small\normalfont}
\definecolor{cai_affil_color}{HTML}{3F8984}

\makeatletter
\renewcommand\AB@affilsepx{\\\protect\Affilfont}
\let\orig@maketitle\maketitle
\renewcommand{\maketitle}{%
  \orig@maketitle%
  \vspace{-1.5em}%
  {\color{cai_primary!30}\hrule height 0.5pt}%
  \vspace{1em}%
}
\makeatother

\title{\LARGE\textcolor{cai_primary}{\textbf{Cybersecurity AI: Hacking Consumer Robots in the AI Era}}}

\author[1,$\dagger$]{Víctor Mayoral-Vilches}
\author[1,$\dagger$]{Unai Ayucar-Carbajo}
\author[$\star$]{Olivier Laflamme}
\author[$\star$]{Ruikai Peng}
\author[1]{María Sanz-Gómez}
\author[1]{Francesco Balassone}
\author[2]{Lucas Apa}
\author[1]{Endika Gil-Uriarte}

\affil[ ]{{\normalfont\textsuperscript{1}\textcolor{cai_primary}{\textbf{Alias Robotics}}, Spain~~~$\vert$~~~\textsuperscript{$\star$}Alias Robotics External Collaborator~~~$\vert$~~~\textsuperscript{2}\textcolor{cai_primary}{\textbf{IOActive}}, USA~~~}}

\makeatletter
\renewcommand\AB@affilnote[1]{}
\makeatother

\affil[*]{
    {\normalfont{\faGithub}~{\tt\footnotesize \href{https://github.com/aliasrobotics/cai}{https://github.com/aliasrobotics/cai}}~~~{\tt\footnotesize\textcolor{cai_primary}{\faEnvelope}~research@aliasrobotics.com}}
}

\makeatletter
\renewenvironment{abstract}{%
  \small
  \noindent\ignorespaces
}{%
  \par
}
\makeatother

\begin{document}

\pagestyle{fancy}
\fancyhf{}
\fancyhead[L]{\textit{\leftmark}}
\renewcommand{\sectionmark}[1]{\markboth{#1}{}}

\date{}

\begingroup
\footnotesize
\setlength{\tabcolsep}{4pt}%
\renewcommand{\arraystretch}{1.2}%
\settowidth{\tablewidth}{%
  \begin{tabular}{@{}lcccccc@{}}
    \textbf{Robot Type} & \textbf{Origin} & \textbf{Category} & \textbf{Attack Surface} & \textbf{Vulns} & \textbf{Assessment Time$^*$} & \textbf{Impact} \\
    Hookii Neomow & China & Outdoor & WiFi, MQTT, ADB, ROS~2 & \textcolor{cai_primary}{\textbf{9}} & $\sim$2.5 hours & Physical + Privacy \\
    Hypershell X & China & Wearable & BLE, REST API, CAN Bus & \textcolor{cai_primary}{\textbf{12}} & $\sim$4 hours & Safety-critical \\
    HOBOT S7 Pro & Taiwan & Indoor & BLE, Cloud API, HTTP & \textcolor{cai_primary}{\textbf{17}} & $\sim$3 hours & Property + Privacy \\
  \end{tabular}%
}%
\global\tablewidth=\tablewidth
\endgroup

\twocolumn[
\maketitle
\vspace{0.5em}

\begin{abstract}
\noindent \textbf{Is robot cybersecurity broken by AI?} Consumer robots---from autonomous lawnmowers to powered exoskeletons and window cleaners---are rapidly entering homes and workplaces, yet their security remains rooted in assumptions of specialized attacker expertise. This paper presents evidence that Generative AI has fundamentally disrupted robot cybersecurity: what historically required deep knowledge of ROS, ROS~2, and robotic system internals can now be automated by anyone with access to state-of-the-art GenAI tools spearheaded by the open source CAI (Cybersecurity AI). We provide empirical evidence through three case studies: (1) compromising a Hookii autonomous lawnmower robot, uncovering fleet-wide vulnerabilities and data protection violations affecting 267+ connected devices, (2) exploiting a Hypershell powered exoskeleton, demonstrating safety-critical motor control weaknesses and credential exposure including access to over 3{,}300 internal support emails, and (3) breaching a HOBOT S7 Pro window cleaning robot, achieving unauthenticated BLE command injection and OTA firmware exploitation. Across these platforms, CAI discovered in an automated manner \textbf{38 vulnerabilities} that would have previously required months of specialized security research. Our findings reveal a stark asymmetry: while offensive capabilities have been democratized through AI, defensive measures often remain lagging behind. We argue that traditional defense-in-depth architectures like the Robot Immune System (RIS) must evolve toward GenAI-native defensive agents capable of matching the speed and adaptability of AI-powered attacks.
\end{abstract}

\vspace{2.5em}
  \begingroup
  \footnotesize
  \setlength{\tabcolsep}{4pt}
  \renewcommand{\arraystretch}{1.2}
  \begin{center}
    \robotimage
    \arrayrulecolor{cai_primary!60}
    \begin{tabular}{@{}lcccccc@{}}
      \toprule
      \rowcolor{cai_primary!12}
      \textbf{Robot Type} & \textbf{Origin} & \textbf{Category} & \textbf{Attack Surface} & \textbf{Vulns} & \textbf{Assessment Time} & \textbf{Impact} \\
      \midrule
      Hookii Neomow & China & Outdoor & WiFi, MQTT, ADB, ROS~2 & \textcolor{cai_primary}{\textbf{9}} & 2.5 hours & Physical + Privacy \\
      Hypershell X & China & Wearable & BLE, REST API, CAN Bus & \textcolor{cai_primary}{\textbf{12}} & 1.5 hours$^\dagger$ & Safety-critical \\
      HOBOT S7 Pro & Taiwan & Indoor & BLE, Cloud API, HTTP & \textcolor{cai_primary}{\textbf{17}} & 3 hours & Property + Privacy \\
      \bottomrule
    \end{tabular}
    \arrayrulecolor{black}
    \captionof{table}{Summary of AI-assisted security assessments on consumer robots. CAI autonomously discovered 38 vulnerabilities across three diverse robotic platforms. Assessment times reflects the time taken by CAI to validate most of the vulnerabilities. $^\dagger$For comparison, an independent external team conducted a human-led assessment of the Hypershell X obtaining similar results; the human-led exercise lasted approximately 5 hours. Unitree G1 is discussed in \cite{mayoral2025humanoid, mayoral2025cybersecurityshort}. A privacy analysis is provided in Appendix~\ref{sec:appendix_privacy} and a complete vulnerability inventory in Appendix~\ref{sec:appendix_vulns}.}
    \label{tab:robot_summary}
  \end{center}
  \endgroup
]

\renewcommand{\thefootnote}{$\dagger$}
\setcounter{footnote}{0}
\footnotetext{These authors contributed equally to this work.}
\renewcommand{\thefootnote}{\arabic{footnote}}

\clearpage

\section{Introduction}

Consumer robotics has entered a new era. Autonomous lawnmowers maintain suburban lawns, powered exoskeletons augment human mobility, and window-cleaning robots scale building facades---all increasingly connected, increasingly autonomous, and increasingly vulnerable. The cybersecurity of these systems has traditionally relied on an implicit assumption: attacking robots requires specialized expertise in robotic middleware (ROS, ROS~2), embedded systems, and cyber-physical system dynamics. Defense-in-depth approaches like the Robot Security Framework (RSF)~\cite{mayoral2018rsf} and Robot Immune System (RIS)~\cite{aliasrobotics2025ris} were developed under this assumption. That assumption is now obsolete, as AI-powered offensive capabilities have fundamentally changed the threat landscape.

This paper presents evidence that Generative AI has fundamentally altered the security model of consumer robotics. Using CAI (Cybersecurity AI)~\cite{cai2025github}, an open-source framework for autonomous security assessment with state-of-the-art performance across diverse security challenges~\cite{sanzgomez2025cybersecurityaibenchmarkcaibench}, we demonstrate that vulnerabilities which previously required months of specialized research can now be discovered and exploited in hours, by anyone with access to state-of-the-art GenAI tools.

Our contributions are twofold:
\begin{enumerate}
\item \textbf{Empirical Evidence of AI-Democratized Robot Hacking}: We present three case studies demonstrating CAI's ability to autonomously compromise consumer robots across different categories: outdoor (Hookii lawnmower), wearable (Hypershell exoskeleton), and indoor (HOBOT S7 Pro window cleaner).


\item \textbf{A Call for GenAI-Native Robot Defense}: We examine how AI has asymmetrically empowered attackers while defensive measures remain largely unchanged. We argue that traditional architectures like the Robot Immune System (RIS)~\cite{aliasrobotics2025ris} must evolve toward AI-native defensive agents.
\end{enumerate}

\subsection{Related Work}\label{sec:related}

\textbf{Robot Cybersecurity.}
The Robot Security Framework (RSF)~\cite{mayoral2018rsf} introduced a standardized four-layer methodology (physical, network, firmware, application) for conducting security assessments in robotics, while the Robot Hacking Manual~\cite{mayoral2022rhm} documented over 100 security flaws and 17 CVE IDs across multiple robotic platforms. The monograph by Zhu et al.~\cite{zhu2021cybersecurity} formalized robot cybersecurity using game-theoretic quantitative modeling, and subsequent reviews~\cite{mayoral2022review} showed that vendors often leave zero-day exposures unpatched for years due to the complexity and cost of securing robotic systems. An early demonstration of real-world consequences was Akerbeltz~\cite{mayoral2020akerbeltz}, the first documented case of industrial robot ransomware targeting Universal Robots platforms. The doctoral thesis \emph{Offensive Robot Cybersecurity}~\cite{mayoral2025offensive} argued that AI was already ``breaking'' traditional robot security models, demonstrating that ``securing robots by hacking them first'' via machine learning and game theory could streamline vulnerability identification and exploitation---foreshadowing the empirical results presented in the current paper. Recent assessments of humanoid robots~\cite{mayoral2025humanoid} have confirmed that even next-generation platforms exhibit the same systemic weaknesses (hardcoded credentials, unencrypted telemetry, unsigned firmware) documented in earlier consumer robots.

Broader surveys corroborate these findings. Botta et al.~\cite{botta2023cybersecurity} catalogue attack vectors and penetration testing tools (ROSPenTo, ROSchaos, ROSploit) for ROS-based systems. Neupane et al.~\cite{neupane2024security} extend the taxonomy across attack surfaces, ethical/legal concerns, and human-robot interaction security, while Haskard and Herath~\cite{haskard2025secure} introduce the concept of ``secure robotics'' at the nexus of safety, trust, and cybersecurity. At the middleware level, Mayoral-Vilches et al.~\cite{mayoral2022sros2} proposed SROS2, a popular series of tools to secure ROS~2 middleware; however, subsequent work has exposed fundamental limitations: Deng et al.~\cite{deng2022insecurity} demonstrated that SROS2 contains architectural vulnerabilities, Sakib et al.~\cite{sakib2025supplychain} showed that supply chain attacks on SROS~2 can compromise autonomous vehicle control via keystore exfiltration, and Soriano-Salvador et al.~\cite{soriano2024rips} proposed RIPS, an intrusion prevention system specifically designed for ROS~2. These studies establish the breadth of robotic attack surfaces; our work demonstrates that AI agents can now autonomously replicate and extend such findings without requiring the years of domain expertise historically necessary.

\textbf{Consumer Robot and IoT Privacy.}
The privacy risks of consumer robots have been documented in targeted studies. Ulsmaag et al.~\cite{ulsmaag2024privacy} showed that passive network eavesdropping of robot vacuum cleaners can reveal user activity patterns from unencrypted header metadata alone. Giese and Luedtke~\cite{giese2024ecovacs} demonstrated at DEF~CON~32 that Ecovacs home robots could be remotely hijacked via Bluetooth to activate cameras and microphones for surveillance. The MQTT protocol, widely used in IoT robotics, presents systemic risks: Laghari et al.~\cite{laghari2024mqtt} provide a comprehensive taxonomy of MQTT ecosystem vulnerabilities, including authentication bypass, topic hijacking, and plaintext data exfiltration---all of which our Hookii assessment confirmed in a production environment. Our work extends these findings by showing that AI agents can autonomously discover such privacy violations alongside technical vulnerabilities.

\textbf{LLM-Assisted Penetration Testing.}
The application of Large Language Models to offensive security has accelerated rapidly. Happe and Cito~\cite{happe2023pwned} were among the first to demonstrate LLM-driven penetration testing, implementing a closed-feedback loop between GPT-3.5-generated actions and vulnerable virtual machines. PentestGPT~\cite{deng2024pentestgpt} formalized this approach with a three-module architecture achieving a 228.6\% task-completion increase over baselines, earning the Distinguished Artifact Award at USENIX Security~2024. Subsequent systems have expanded the paradigm: AutoAttacker~\cite{xu2024autoattacker} automates post-breach lateral movement using LLM-guided Metasploit campaigns, while HackSynth~\cite{muzsai2024hacksynth} introduces a dual-module Planner/Summarizer architecture evaluated against 200 CTF challenges. Mayoral-Vilches et al.~\cite{mayoral2026superintelligence} trace the full arc from PentestGPT through CAI, proposing a trajectory toward ``cybersecurity superintelligence'' where AI exceeds human capability in both speed and strategic reasoning. CAI~\cite{aliasrobotics2025cai} builds on this lineage but uniquely targets real-world robotic systems rather than CTF environments or simulated networks, requiring autonomous handling of heterogeneous protocols (BLE, MQTT, ROS~2, REST APIs) and physical system interactions.

\subsection{The End of Security Through Obscurity}

Robot security has historically benefited from obscurity. The specialized knowledge required to understand ROS communication patterns, analyze robotic firmware, or exploit cyber-physical interactions created natural barriers to entry. This expertise barrier is collapsing. Large Language Models trained on robotics documentation, security research, and exploit databases can now guide attackers through complex robotic systems without requiring years of specialized training. CAI exemplifies this shift: it combines domain knowledge with autonomous reasoning capabilities that allow it to formulate hypotheses, test attack vectors, and chain vulnerabilities---tasks that previously demanded human experts.

Our assessments revealed not only technical vulnerabilities but also systemic failures in data governance. Two of the three robots assessed exhibited confirmed GDPR compliance failures, including data transmission without documented user consent, absence of data subject rights mechanisms, and cross-border data transfers without identified legal basis. CAI discovered these compliance failures alongside traditional security vulnerabilities, demonstrating that AI-powered assessments naturally extend beyond pure exploitation into regulatory risk identification.

\section{Methodology}\label{sec:methodology}

We evaluated CAI's capability to assess consumer robot security across three representative platforms:

\textbf{Target 1 -- Hookii Neomow (Lawnmower Robot)}: An autonomous lawn mower running Debian~11 with ROS~2 Humble on Linux 5.10.110. Features WiFi connectivity, MQTT-based cloud telemetry to AWS infrastructure, an HD camera (1280$\times$720), Livox LiDAR for 3D mapping, GPS/IMU sensors, and a 4G/LTE cellular modem providing an independent data channel.

\textbf{Target 2 -- Hypershell X (Powered Exoskeleton)}: A dual-motor assistive walking exoskeleton with an ESP32 masterboard communicating with two STM32 motor controllers via CAN Bus. Uses BLE~4.0+ via Nordic UART Service for mobile app control, REST API endpoints for cloud services (hosted on both Chinese and international servers), and OTA firmware updates.

\textbf{Target 3 -- HOBOT S7 Pro (Window Cleaning Robot)}: A window cleaning robot built on a Silicon Labs EFR32BG22 SoC with BLE connectivity. The robot communicates locally via BLE and reaches the Gizwits IoT cloud platform indirectly through a companion smartphone app for remote control and telemetry (34 data attributes). OTA updates are delivered via both BLE and plaintext HTTP channels from \texttt{hobot.com.tw}.

Each assessment followed a standardized protocol using CAI~\cite{cai2025github}, a CLI-based cybersecurity agent, with a human operator in the loop: (1) CAI was provided only with the robot's product name---no technical documentation or prior research was collected manually; (2) CAI autonomously discovered network interfaces and wireless attack surfaces; (3) CAI systematically tested for security weaknesses with human oversight; (4) working exploits were developed and validated where feasible; (5) successful exploits were evaluated for real-world impact. Human operators guided the assessment process and intervened for safety reasons when tests extrapolated to the cloud infrastructure of vendors.

Vulnerability severity was assessed by the authors using CVSS~3.1 base metrics. Assessment durations reported in Table~\ref{tab:robot_summary} are approximate and reflect active analysis time; they do not represent controlled measurements.

\section{Results}\label{sec:results}

\begin{figure}[h!]
  \centering
  \begin{tikzpicture}
    \begin{axis}[
      width=0.75\linewidth,
      height=5.5cm,
      xbar,
      xmin=0, xmax=20,
      xlabel={Vulnerabilities Discovered},
      xlabel style={font=\small\bfseries},
      symbolic y coords={HOBOT S7 Pro,Hypershell X,Hookii Neomow},
      ytick=data,
      y dir=reverse,
      bar width=12pt,
      enlarge y limits=0.25,
      yticklabel style={font=\footnotesize},
      tick label style={font=\small},
      nodes near coords,
      every node near coord/.style={font=\small,black},
      axis x line*=bottom,
      axis y line*=left,
      xmajorgrids=true,
      grid style={draw=gray!10,line width=0.3pt},
      legend style={at={(1.1,0.05)}, anchor=south east, legend columns=1, font=\small},
    ]
      \addlegendimage{area legend,fill=cai_primary!85,draw=cai_primary!90}
      \addlegendentry{Critical/High}
      \addlegendimage{area legend,fill=cai_primary!40,draw=cai_primary!50}
      \addlegendentry{Medium/Low}

      \addplot[sharpbarh,fill=cai_primary!85,draw=cai_primary!90,bar shift=-6pt,
        every node near coord/.append style={yshift=-7pt, xshift=2pt}]
        coordinates {(10,HOBOT S7 Pro) (12,Hypershell X) (8,Hookii Neomow)};
      \addplot[sharpbarh,fill=cai_primary!40,draw=cai_primary!50,bar shift=6pt,
        every node near coord/.append style={yshift=7pt, xshift=2pt}]
        coordinates {(7,HOBOT S7 Pro) (0,Hypershell X) (1,Hookii Neomow)};
    \end{axis}
  \end{tikzpicture}
  \caption{Vulnerability severity distribution across consumer robots assessed using CVSS~3.1. CAI identified 30 Critical/High severity vulnerabilities and 8 Medium/Low severity issues across the three platforms. All 12 Hypershell vulnerabilities were assessed as Critical or High, reflecting design-level security deficiencies.}
  \label{fig:vuln_severity}
\end{figure}
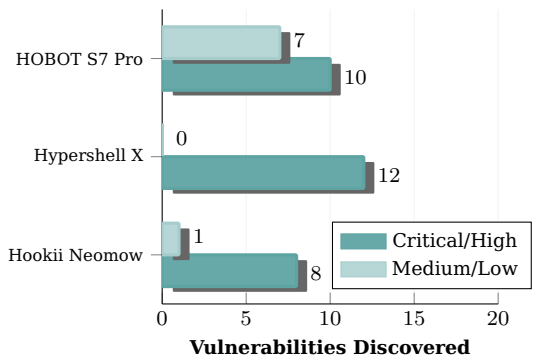

\subsection{Case Study 1: Lawnmower}\label{sec:lawnmower}

CAI's assessment of the Hookii Neomow revealed pervasive security deficiencies affecting the entire Hookii robot fleet. The assessment identified 9 distinct vulnerabilities, including 4 assessed as Critical.

CAI began with network reconnaissance on the local network, immediately identifying an \textbf{unauthenticated Android Debug Bridge (ADB)} service on port~5555 (CVSS~10.0). A simple \texttt{adb connect} command granted unrestricted root access to the robot's Debian~11 system without any authentication. From this foothold, CAI extracted hardcoded MQTT credentials from \texttt{/home/linaro/task/.mqtt.json}---credentials that are \textbf{identical across the entire Hookii fleet} (CVSS~9.8), as confirmed by subsequent broker-side enumeration.

Using these fleet-wide credentials, CAI connected to Hookii's EMQX MQTT broker at \texttt{neomowx.hookii.com} and discovered it was configured with \textbf{default administrative credentials} (\texttt{admin:public}, CVSS~9.1). Through the EMQX management API (port~8081), CAI enumerated \textbf{267 connected robots} with their IP addresses and 333 active MQTT subscriptions, and confirmed the ability to publish commands to any robot in the fleet. The vendor's MySQL~8.0.24 database on port~3306 was publicly accessible without firewall protection (CVSS~9.8), with 11 database users enumerable via authentication response analysis.

\textbf{Data Transmission and Privacy}: CAI identified continuous data transmission via unencrypted MQTT (port~1883, TLS explicitly disabled in the robot's configuration with \texttt{use\_tls:~0}). The robot transmitted GPS coordinates every 30 seconds and operational telemetry every 5 seconds, totalling 8{,}472 packets (566{,}669 bytes) in a single observed session. Local storage contained 8.4GB+ of telemetry logs, 18MB+ of costmap camera images, and 456MB of 3D property mapping data generated by Livox LiDAR---including a 206MB point cloud, property boundary coordinates, and mowing zone definitions. A 4G/LTE cellular modem (carrier: Asia.bics) provided an independent data channel that bypasses local network security controls. Analysis of the EMQX broker confirmed at least one external client had downloaded 724.98MB of data over a 49-day connection period. CAI documented potential violations of multiple GDPR articles (see Appendix~\ref{sec:appendix_privacy}), including the absence of any opt-in or opt-out consent mechanism and data transfers to AWS infrastructure in the United States.

\textbf{Impact}: Chaining the ADB access, fleet-wide credentials, and EMQX default credentials enabled access to all 267+ connected Hookii robots simultaneously, including the ability to publish to command topics and access fleet-wide telemetry.

\subsection{Case Study 2: Mountain Exoskeleton}\label{sec:exoskeleton}

The Hypershell powered exoskeleton presented the highest-stakes target, with vulnerabilities potentially affecting user physical safety. Supported by CAI, we identified 12 vulnerabilities, all assessed as Critical or High severity. Two of these (HS-11, HS-12 in Appendix~\ref{sec:appendix_vulns}) were identified through static analysis of decompiled application code and were not validated with working exploits.

\textbf{Discovery Process}: We began by scanning for BLE advertisements, identifying devices broadcasting as ``hypershell.'' Connection attempts revealed \textbf{no BLE authentication}---any BLE client within range can connect and send arbitrary commands via the Nordic UART Service. CAI discovered a critical \textbf{Insecure Direct Object Reference (IDOR)} chain: device IDs are the reversed bytes of the BLE MAC address (e.g., MAC \texttt{F8:B3:B7:D5:23:16} becomes device ID \texttt{1623d5b7b3f8}). This predictable derivation, combined with unauthenticated API endpoints, allowed CAI to query \texttt{/device/isDeviceBind}, \texttt{/record/deviceRecord}, and \texttt{/query/getUserInfoAndDeviceInfo} to retrieve owner emails, serial numbers, complete usage histories, and battery data for arbitrary devices---confirmed with working proof-of-concept scripts.

Code analysis of the decompiled Flutter application revealed that the BLE ``pairing secret'' is the \texttt{motorVersion} field returned by the server API to any authenticated user querying any device. Further analysis identified \textbf{unsigned OTA firmware updates} protected only by CRC16 checksums, with firmware binaries publicly accessible at predictable URLs on \texttt{api.hypershell.net}.

\textbf{Credential Exposure}: Through reverse engineering, we identified plaintext SMTP credentials for \texttt{it@hypershell.tech} in application heap dumps, providing access to an estimated 3{,}300+ internal support emails containing, among other data, PayPal and Shopify account recovery codes. Root MySQL database credentials for both Chinese and international servers were also identified in application artifacts, though live database access was not tested. Feishu (internal collaboration platform) API tokens were confirmed functional, providing access to a support database containing 64+ customer tickets with user IDs, device IDs, and email addresses.

\textbf{Impact}: The BLE protocol exposes 177 commands (enumerated via decompilation), including motor control commands (\texttt{SET\_MOTOR\_CUSTOM}, \texttt{SET\_MODE}, \texttt{SET\_RESPONSE\_SPEED}). Static analysis of the \texttt{int8ToUint8} conversion function identified a potential integer overflow (CWE-190) that could cause motor parameter inversion; this finding was not confirmed with a working exploit. The combination of unauthenticated BLE access with motor control commands presents a potential physical safety risk to users.

\subsection{Case Study 3: Window Cleaner}\label{sec:window}

The HOBOT S7 Pro window cleaning robot exhibited security deficiencies across its BLE interface and Gizwits cloud platform. CAI identified 17 vulnerabilities---the highest count among the three targets.

\textbf{Discovery Process}: CAI enumerated BLE GATT services and found the device accepts connections \textbf{without pairing, bonding, or authentication of any kind} (CVSS~9.1). All GATT services were immediately accessible with full read/write permissions. Through APK decompilation of the HOBOT mobile application (which lacked code obfuscation), CAI reverse-engineered the complete BLE command protocol:

\vspace{0.3em}
\noindent\texttt{Old format: 53 CC 00 00 00 45 [XOR]}\\
\texttt{New format: 55 AA LL LL [cmd+data] [XOR]}
\vspace{0.3em}

CAI validated 16 distinct commands via PoC including motor control, water spray activation, directional movement, factory reset, and firmware version queries---all executable without authentication. The integrity mechanism is a single XOR byte (CWE-328), trivially reproducible, with no replay protection (CWE-294): identical commands were accepted and processed repeatedly in testing.

\textbf{OTA and Cloud Exploitation}: CAI identified an \textbf{unauthenticated Silicon Labs OTA service} (CVSS~9.8) that accepted arbitrary firmware data writes without cryptographic signature verification (confirmed via PoC with test payloads). Legitimate firmware downloads are served over \textbf{plaintext HTTP} from \texttt{hobot.com.tw}, enabling potential MITM firmware replacement. From the APK, CAI extracted hardcoded Gizwits cloud credentials (\texttt{app\_id}, \texttt{app\_secret}, \texttt{ProductKey}, \texttt{ProductSecret}), enabling anonymous API access. The Gizwits platform exposed a 34-attribute product schema including 18 writable control attributes and real-time position tracking data (\texttt{status\_posture\_batch}: 126 bytes every 0.7 seconds, per the vendor's schema description).

During initial API probing, CAI observed a \textbf{cross-tenant data leakage} on the Gizwits platform: using HOBOT's application credentials, it retrieved the product schema of an unrelated vendor's smart air conditioner, including 7 writable control attributes. This observation was not reproducible in subsequent probing phases, suggesting it may depend on platform caching or load conditions.

\textbf{Impact}: Chaining the unauthenticated BLE access with motor control commands could enable an attacker within BLE range (up to $\sim$70m per the RF-star BG22A1 module datasheet) to disable suction motors while the robot is attached to a window. The unauthenticated OTA service accepts firmware data writes, as confirmed by PoC testing with non-destructive payloads.

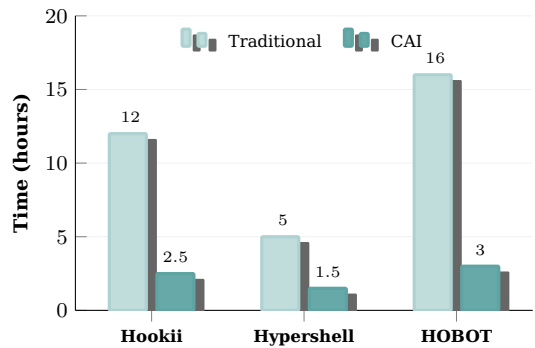
\begin{figure}[h!]
  \centering
  \begin{tikzpicture}
    \begin{axis}[
      width=0.95\columnwidth,
      height=5.5cm,
      ybar=4pt,
      bar width=14pt,
      ymin=0, ymax=20,
      enlarge x limits=0.25,
      symbolic x coords={Hookii,Hypershell,HOBOT},
      xtick=data,
      xticklabel style={font=\footnotesize\bfseries},
      ylabel={Time (hours)},
      ylabel style={font=\small\bfseries},
      tick label style={font=\small},
      ymajorgrids=true,
      grid style={draw=gray!10,line width=0.3pt},
      nodes near coords,
      every node near coord/.style={font=\scriptsize,anchor=south},
      legend style={
        at={(0.5,0.98)},
        anchor=north,
        font=\footnotesize,
        draw=none,
        fill=none,
        column sep=6pt,
      },
      legend columns=2,
      axis x line*=bottom,
      axis y line*=left,
    ]
      \addplot[sharpbar,fill=cai_primary!35,draw=cai_primary!45,
        every node near coord/.append style={yshift=1pt, xshift=-8pt}]
        coordinates {(Hookii,12) (Hypershell,5) (HOBOT,16)};
      \addlegendentry{Traditional}
      \addplot[sharpbar,fill=cai_primary!85,draw=cai_primary!90,
        every node near coord/.append style={yshift=1pt, xshift=9pt}]
        coordinates {(Hookii,2.5) (Hypershell,1.5) (HOBOT,3)};
      \addlegendentry{CAI}
    \end{axis}
  \end{tikzpicture}
  \caption{Assessment time comparison between traditional expert-led and CAI-led security assessments across the three robot platforms (lower is better). CAI reduced assessment time by 3--5$\times$ across all platforms.}
  \label{fig:time_comparison}
\end{figure}

\section{Toward GenAI-Native Robot Defense}\label{sec:defense}

Our results demonstrate an urgent need to evolve robot security architectures. RIS provides a foundation, but its static, rule-based approach is not designed to counter AI-powered attacks that autonomously chain vulnerabilities across BLE, cloud APIs, and firmware update channels. We propose three evolutionary directions that build upon one another.

First, GenAI-native defense should employ adaptive threat detection through behavioral AI models that learn normal robot operation patterns, such as detecting anomalous BLE connection patterns or unusual MQTT topic subscriptions. These systems should incorporate adversarial simulation that continuously generates novel attack scenarios, along with context-aware response mechanisms capable of distinguishing between a legitimate firmware update and a malicious OTA injection.

Second, a defensive AI agent should be capable of autonomous patch generation, validating and deploying fixes in real-time---for instance, enabling TLS on MQTT connections, rotating hardcoded credentials, or disabling exposed debug interfaces such as the ADB service identified on Hookii robots---while maintaining system functionality during remediation.

Third, consumer robots should participate in coordinated defense networks through AI agent to agent and fleet-wide threat intelligence sharing. The Hookii fleet-wide credential reuse demonstrates why per-device credentials are essential, and distributed anomaly detection leveraging fleet-wide visibility would enable collective response to discovered attack vectors.

\section{Discussion}\label{sec:discussion}

Our findings carry implications across multiple dimensions.

\textbf{Manufacturers} can no longer rely on the assumption that specialized knowledge requirements provide adequate security. All three assessed robots exhibited fundamental authentication failures---unauthenticated ADB (Hookii), unauthenticated BLE commands (Hypershell, HOBOT), and default administrative credentials (Hookii EMQX). These do not represent sophisticated attack vectors; they reflect an absence of security consideration during product development. The finding that a single set of hardcoded MQTT credentials provides access to 267+ Hookii robots underscores the risk of fleet-wide credential reuse.



\textbf{Vulnerability disclosure, international cooperation, and the CVE system in the AI era.} Our results expose a compounding crisis across the entire vulnerability management pipeline---from discovery through disclosure to remediation---that is particularly acute in consumer robotics.

On the discovery side, AI has fundamentally changed the economics. In approximately 7~hours of cumulative assessment time---parallelizable to roughly 3~hours with three concurrent agents---CAI helped identify 38 vulnerabilities across three robots, the majority of which would qualify for unique CVE identifiers. By comparison, traditional expert-led assessments of the same platforms required an estimated 33 effective hours (Figure~\ref{fig:time_comparison}), and when accounting for the multiple security specialists involved, the human effort translates to person-weeks of professional time. \emph{What a multidisciplinary security team accomplishes over the course of a working week, three AI agents can now replicate in an afternoon}. Extrapolating this rate across the consumer robotics market suggests that AI-powered assessments could generate vulnerability reports at a pace that overwhelms existing infrastructure. This is not hypothetical. The NIST National Vulnerability Database (NVD) entered crisis in 2024, with 93.4\% of newly submitted CVEs remaining unanalyzed for months~\cite{nvdbacklog2024}. MITRE's CVE program itself nearly collapsed when DHS allowed its funding contract to lapse in April 2025~\cite{krebs2025cvefunding}, prompting the creation of a CVE Foundation and the European GCVE decentralized alternative~\cite{gcve2026}. Meanwhile, AI systems are accelerating the other side of the equation: Google's Big Sleep agent discovered a previously unknown zero-day in SQLite that 150 CPU-hours of traditional fuzzing had missed~\cite{google2024bigsleep}, and recent work has shown that LLM-based frameworks can automatically reproduce working exploits for published CVEs at an average cost of \$2.77 per vulnerability. The vulnerability management paradigm---static databases, periodic CVSS scoring, manual NVD enrichment---was designed for an era of dozens of disclosures per week, not the hundreds per day that AI-accelerated discovery is producing.

On the disclosure and remediation side, the situation is equally strained---and in robotics, compounded by a geographic dimension. The robots assessed in this study originate from manufacturers in China and Taiwan (Table~\ref{tab:robot_summary}). This reflects a structural reality: East Asian manufacturers---and Chinese firms in particular---have emerged as world leaders in consumer robot hardware, producing platforms of remarkable mechanical and electromechanical sophistication. The quality of the hardware engineering in all three assessed platforms was evident throughout our analysis. However, the cybersecurity maturity of these products does not match the caliber of their hardware design, and the channels for coordinated vulnerability disclosure remain underdeveloped. Following our assessments, we attempted to contact all three manufacturers to report the identified vulnerabilities. Hypershell explicitly declined to engage, responding: \emph{``At this time, Hypershell is not pursuing vulnerability disclosure reports or external security research submissions.''} This is consistent with a broader pattern observed across East Asian robotics and IoT manufacturers~\cite{mayoral2025humanoid}, where security reports from international researchers are either ignored or declined, sometimes citing internal policies or regulatory constraints on engagement with external security researchers.

The result is a pipeline that is broken at every stage: AI discovers vulnerabilities faster than CVE infrastructure can catalogue them, and manufacturers decline to remediate them even when notified. In robotics, unlike in traditional software, this is not merely a data security concern. An exoskeleton with exploitable motor control commands, a lawnmower with fleet-wide remote access, or a window cleaner whose suction can be disabled mid-operation present risks that enter the domain of bodily harm. Moreover, robotic vulnerabilities span BLE protocols, MQTT configurations, ROS~2 middleware, REST APIs, OTA update mechanisms, and cloud platforms simultaneously---each requiring distinct remediation workflows that the CVE system was never designed to coordinate. Consistent with this analysis, we have chosen not to apply for CVE identifiers for the 38 vulnerabilities reported in this study. In our assessment, the CVE system in its current state does not meaningfully contribute to improving the security posture of affected products; it primarily serves as a credentialing mechanism within the security community rather than as a driver of actual remediation.

We argue that two parallel reforms are necessary. First, the vulnerability management paradigm must evolve from static databases and periodic scoring toward real-time, context-aware, and AI-augmented triage systems capable of matching the pace at which vulnerabilities are now discovered. Second, the international robotics community---and East Asian manufacturers specifically, given their dominant and growing market position---must establish effective channels for security research cooperation. 
The alternative---a growing fleet of globally deployed robots with known, unpatched, safety-relevant vulnerabilities discovered faster than any existing system can process---is untenable.

We acknowledge the dual-use nature of this research~\cite{moderndiplomacy2025}. All vulnerabilities were reported to manufacturers before publication, and specific exploitation details are withheld pending patches.

\textbf{Limitations}: Three robot platforms cannot represent the entire consumer robotics market. Lab conditions may not capture all real-world scenarios. CAI's effectiveness may vary across different architectures and communication protocols. Long-term attack persistence was not evaluated. Two Hypershell vulnerabilities (HS-11, HS-12) were identified through static analysis and lack working proof-of-concept exploits. The Gizwits cross-tenant data leakage (HB-15) was observed once and was not reproducible in subsequent attempts.

\section{Conclusion}\label{sec:conclusion}

This paper presents evidence that Generative AI has fundamentally changed the threat landscape for consumer robotics. CAI's ability to autonomously discover and exploit vulnerabilities across diverse robot platforms---a Hookii lawnmower (9 vulnerabilities, fleet-wide access to 267+ devices), a Hypershell exoskeleton (12 vulnerabilities, all Critical/High, with credential exposure affecting internal support infrastructure), and a HOBOT S7 Pro window cleaner (17 vulnerabilities spanning BLE, cloud, and firmware)---indicates that the expertise barrier historically protecting robot security has eroded substantially.

The 38 vulnerabilities identified include unauthenticated root access (CVSS~10.0), fleet-wide credential reuse enabling mass device access, unsigned firmware updates, safety-relevant motor control exposure, and GDPR compliance failures spanning multiple regulatory articles. None of these required prior robotic expertise to exploit once CAI identified them.

The implications are clear: \textbf{robot cybersecurity must be rearchitected for the AI era}. Traditional defense-in-depth architectures are insufficient against AI-powered offensive capabilities. The industry must respond: manufacturers should conduct AI-assisted security assessments before deployment; architectures like RIS~\cite{aliasrobotics2025ris} must evolve toward GenAI-native defensive agents; the robotics industry needs coordinated threat intelligence sharing; and regulators must mandate security and privacy compliance for connected consumer robots. That the assessed platforms originate from manufacturers in China and Taiwan---reflecting East Asia's dominance in consumer robot hardware---underscores both the market's geographic concentration and the urgency of establishing effective international channels for coordinated vulnerability disclosure and remediation in robotics. The democratization of robot security through AI is not a future threat---it is present reality.

\section{Acknowledgements}

This research was partly funded by the European Innovation Council (EIC) accelerator project ``RIS'' (GA 101161136).

\bibliography{bibliography}


\clearpage
\onecolumn
\appendix

\section{Privacy Analysis}\label{sec:appendix_privacy}

Beyond traditional cybersecurity vulnerabilities, CAI's assessments revealed privacy and data protection concerns across all three consumer robot platforms. This appendix documents the observed data collection practices and the status of consent and data subject rights mechanisms. All findings reported below are based on direct observation of device behavior, configuration file analysis, network traffic inspection, and API endpoint probing conducted during the assessments.

\subsection{Data Collection Practices}\label{sec:data_collection}

All three robots were observed to collect and transmit personal data in ways that appear to exceed what is strictly necessary for their primary operational function.

\subsubsection*{Hookii Neomow}

Six categories of data collection were confirmed through passive analysis of device logs, configuration files, and network traffic:

\textbf{(1) MQTT Telemetry.} The robot maintains a persistent connection from \texttt{robot\_base\_node} (PID~5179) to the vendor's EMQX broker at \texttt{3.239.235.126:1883}. During a single observed session, 8{,}472 packets totalling 566{,}669 bytes were transmitted. The connection transmits heartbeat messages every 5 seconds and status updates (including GPS coordinates) every 30 seconds. TLS is explicitly disabled in the robot's configuration (\texttt{use\_tls:~0}), and all data---including location coordinates---is transmitted in plaintext. Local MQTT logs total 24MB active plus 50MB rotated. Separate analysis of the EMQX broker (via the default \texttt{admin:public} credentials) revealed that at least one external client (IP: 193.77.49.110, Slovenian ISP) had maintained a 49-day connection and downloaded 724.98MB of data at an 11.3:1 download/upload ratio.

\textbf{(2) File Uploads.} The robot performs automated HTTP uploads to \texttt{neomowx.app.hookii.com:10752} via \texttt{/api/v1/common/upload/device/file}. Log analysis revealed 20+ consecutive upload attempts in a single session at approximately 2-second intervals (timestamps: 2025-12-29 08:36:27 through 08:37:06, extracted from \texttt{/userdata/log/robot\_base.log}).

\textbf{(3) Camera Images.} An HD camera (1280$\times$720 at 10~fps, model: WAW\_RGB\_Camera, process: \texttt{hookii\_usb\_cam\_node\_exe}) captures images stored in \texttt{/userdata/log/costmap\_image/} as timestamped \texttt{tar.gz} archives totalling 18MB+. Archives contain 10+ sequential frames each, triggered by operational events such as navigation failures.

\textbf{(4) Property Mapping.} The robot generates 456MB+ of 3D property mapping data via Livox LiDAR, stored in \texttt{/userdata/HOOKII\_MAP/}. This includes a 206MB point cloud file (\texttt{.pcd}), property boundary coordinates (\texttt{boundary.csv}), mowing zone definitions (\texttt{mow\_area.csv}), navigation paths (\texttt{lane.csv}), and charging station location (\texttt{station.csv}), distributed across 3{,}213 subdirectories.

\textbf{(5) Cellular Data Channel.} A 4G/LTE modem (\texttt{/dev/usb-4g\_1}, carrier: Asia.bics) provides an independent internet connection via PPP. This channel operates independently of the user's WiFi network and cannot be disabled through any documented user-facing mechanism.

\textbf{(6) WiFi Credential Storage.} User WiFi credentials are stored in plaintext in \texttt{/etc/wpa\_supplicant.conf}. A default access point password (\texttt{neomowx123456}) is hardcoded in \texttt{/home/linaro/.wpa\_conf.json} and appears to be common across all devices.

Total local data storage observed on the assessed device exceeded 8.4GB, including telemetry logs, mapping data, camera images, and coredumps.

\subsubsection*{Hypershell X}

The Hypershell exoskeleton collects mobility and activity data including step counts, altitude changes, and walking patterns for each usage session. This data is stored server-side and was found to be accessible via unauthenticated API endpoints using only the device ID (derivable from passively observed BLE advertisements, as described in Section~\ref{sec:exoskeleton}).

The credential exposure identified during the assessment provides access to several repositories of personal data:
\begin{itemize}
\item \textbf{Email system}: SMTP/IMAP credentials for \texttt{it@hypershell.tech} (identified in application heap dumps) provide access to the vendor's support mailbox. The email retrieval script fetched the 20 most recent messages; the total mailbox size was estimated at approximately 3{,}300 emails based on IMAP mailbox metadata, though this figure was not verified by full enumeration.
\item \textbf{Support database}: Confirmed-functional Feishu API tokens provided access to a support ticket database containing 64+ entries with user IDs, device IDs, email addresses, debug log URLs, and device photographs.
\item \textbf{API endpoints}: Both \texttt{api.hypershell.net:9889} (international) and \texttt{api.hypershell.cn:9890} (China) expose user profiles and device-to-owner mappings via IDOR-vulnerable endpoints, confirmed with proof-of-concept scripts.
\end{itemize}

\subsubsection*{HOBOT S7 Pro}

The Gizwits IoT cloud platform schema extracted during the assessment defines 34 data attributes (18 writable, 16 read-only) transmitted between the device and cloud. Privacy-relevant attributes include:
\begin{itemize}
\item \texttt{status\_posture\_batch}: 126 bytes transmitted every 0.7 seconds (per the vendor's Chinese-language schema description, translated: ``device sends 126 bytes every 0.7s, each containing 7 entries of 18 bytes''), representing real-time position data.
\item \texttt{initial\_map\_config}: Floor plan dimensions (columns, rows, pitch in mm).
\item \texttt{status\_docker\_coordnate}: Charging dock coordinates (fixed location within the residence).
\item \texttt{status\_total\_work\_hour}: Cumulative operating time up to 99{,}999{,}999 seconds ($\sim$3.2 years).
\item \texttt{mode\_scheduling}: 7-day cleaning schedule (28 bytes encoding day, hour, minute, and mode).
\item \texttt{status\_hosting}: An online presence indicator (4 bytes, transmitted periodically).
\end{itemize}

Data flows to Gizwits endpoints across three regions (US: \texttt{usapi.gizwits.com}, CN: \texttt{api.gizwits.com}, EU: \texttt{euapi.gizwits.com}). The APK was found to use hardcoded DNS resolver 114.114.114.114 (a Chinese public DNS service). Regional data isolation was not assessed.

\subsection{Consent Mechanisms and Data Subject Rights}\label{sec:consent}

None of the three robots were found to implement consent management or data subject rights mechanisms consistent with GDPR requirements.

\subsubsection*{Hookii Neomow}

The Hookii robot provides \textbf{no observable opt-in or opt-out mechanism} for data collection. Telemetry transmission begins automatically upon device power-on and continues without interruption. No in-app or on-device control was identified to limit data collection scope, frequency, or recipients. No transparency notices were identified informing users of the extent of data collection. The GDPR violation report generated during the assessment cited violations across Articles~5(1)(a), 5(1)(b), 5(1)(c), 5(1)(e), 5(1)(f), 6, 7, 13, 14, 15, 16, 17, 18, 20, 21, 25, 32, 33, 35, and 44--49---a total of 21 articles.

Data is transmitted to AWS infrastructure in the United States (3.239.235.126) and a test server was identified in China (14.18.190.219, \texttt{test.hookii.com}). No Standard Contractual Clauses, adequacy decisions, or other transfer mechanisms under GDPR Articles~44--49 were identified.

\subsubsection*{Hypershell X}

No consent management interface was identified in the Hypershell mobile application or API. Users cannot request, export, or delete their activity history through any documented mechanism. The dual infrastructure in China (\texttt{api.hypershell.cn:9890}) and internationally (\texttt{api.hypershell.net:9889}) means user data is accessible from both endpoints without geographic restriction. The persistence of plaintext credentials (SMTP, database, Feishu API) in production application artifacts constitutes a de facto data exposure, potentially triggering breach notification obligations under GDPR Article~33.

\subsubsection*{HOBOT S7 Pro}

CAI conducted a systematic probe of 18 GDPR-related API endpoint patterns across the Gizwits cloud platform. All requests returned HTTP~404:

\vspace{0.5em}
\begin{small}
\begin{center}
\setlength{\tabcolsep}{4pt}
\renewcommand{\arraystretch}{1.1}
\arrayrulecolor{cai_primary!60}
\begin{tabular}{@{}llc@{}}
\toprule
\rowcolor{cai_primary!12}
\textbf{GDPR Right} & \textbf{Endpoints Tested} & \textbf{Found} \\
\midrule
Art.~15 -- Right of Access & \texttt{/app/users/me/data}, \texttt{/app/gdpr/access}, +4 & 0/6 \\
Art.~17 -- Right to Erasure & \texttt{/app/users/me/delete}, \texttt{/app/gdpr/erasure}, +3 & 0/5 \\
Art.~20 -- Data Portability & \texttt{/app/users/me/export}, \texttt{/app/gdpr/portability}, +2 & 0/4 \\
Consent Management & \texttt{/app/consent}, \texttt{/app/privacy/consent}, +1 & 0/3 \\
\bottomrule
\end{tabular}
\end{center}
\end{small}
\vspace{0.5em}

\noindent While \texttt{DELETE /app/users} returns HTTP~204 and invalidates the session token, the assessment found no confirmation of backend data erasure, no data export option offered before deletion, no cooling-off period, and no email confirmation.

Data flows to Gizwits cloud endpoints across three regions. Under GDPR Articles~44--49, transfers to countries without an adequacy decision require appropriate safeguards; none were identified during the assessment.

\clearpage
\section{Complete Vulnerability Inventory}\label{sec:appendix_vulns}

Table~\ref{tab:vuln_inventory} provides the full inventory of all 38 vulnerabilities identified by CAI across the three consumer robot platforms. Vulnerabilities are grouped by target and ordered by severity. CVSS scores follow version~3.1 base metrics and were assessed by the authors based on observed impact and exploitability. Entries marked N/A under CVSS represent regulatory compliance findings rather than traditional software vulnerabilities. Findings marked with $\dagger$ were identified through static analysis of decompiled code and lack working proof-of-concept exploits. The finding marked with $\ddagger$ was observed once during API probing and was not reproducible in subsequent attempts.

\vspace{1em}

\begin{small}
\setlength{\LTpre}{0pt}
\setlength{\LTpost}{0pt}
\renewcommand{\arraystretch}{1.15}
\arrayrulecolor{cai_primary!60}
\begin{longtable}{@{}l l p{5.8cm} l r l@{}}
\caption{Complete vulnerability inventory across all assessed consumer robots. 38 vulnerabilities total: 16 Critical, 14 High, 6 Medium, 2 Low.}\label{tab:vuln_inventory}\\
\toprule
\rowcolor{cai_primary!12}
\textbf{ID} & \textbf{Robot} & \textbf{Vulnerability} & \textbf{CWE} & \textbf{CVSS} & \textbf{Severity} \\
\midrule
\endfirsthead
\multicolumn{6}{l}{\small\textit{Table~\ref{tab:vuln_inventory} continued from previous page}}\\[0.5em]
\toprule
\rowcolor{cai_primary!12}
\textbf{ID} & \textbf{Robot} & \textbf{Vulnerability} & \textbf{CWE} & \textbf{CVSS} & \textbf{Severity} \\
\midrule
\endhead
\midrule
\multicolumn{6}{r}{\small\textit{Continued on next page}}\\
\endfoot
\bottomrule
\endlastfoot

\rowcolor{cai_light}
\multicolumn{6}{l}{\textbf{\textcolor{cai_primary}{Hookii Neomow --- Autonomous Lawnmower Robot}}} \\
\midrule

HK-01 & Hookii & Unauthenticated ADB service on port 5555 granting unrestricted root shell access without any authentication & CWE-306 & 10.0 & Critical \\

HK-02 & Hookii & Hardcoded fleet-wide MQTT credentials identical across all Hookii robots, stored in plaintext in \texttt{.mqtt.json} & CWE-798 & 9.8 & Critical \\

HK-03 & Hookii & Vendor MySQL~8.0.24 database (port 3306) publicly accessible; 11 users enumerable via authentication response analysis & CWE-284 & 9.8 & Critical \\

HK-04 & Hookii & EMQX MQTT broker management API (port 8081) accessible with default credentials (\texttt{admin:public}); enables fleet enumeration and command publication & CWE-1392 & 9.1 & Critical \\

HK-05 & Hookii & Continuous data transmission via unencrypted MQTT: GPS coordinates, camera images, 456MB property maps, 8.4GB+ telemetry logs & CWE-319 & 8.1 & High \\

HK-06 & Hookii & Unencrypted MQTT communications (TLS explicitly disabled with \texttt{use\_tls:~0}); all telemetry and commands in plaintext on port 1883 & CWE-319 & 8.1 & High \\

HK-07 & Hookii & Independent 4G/LTE cellular modem provides data channel bypassing local network security controls; not user-disableable & CWE-923 & 7.5 & High \\

HK-08 & Hookii & Outdated OpenSSH 8.4p1 and PolicyKit 0.105 (CVE-2021-4034 PwnKit) present on device & CWE-1104 & 7.2 & High \\

HK-09 & Hookii & World-writable critical system files (\texttt{/etc/ppp/ip-up}, \texttt{/etc/fstab}, package source lists) & CWE-732 & 6.5 & Medium \\

\midrule
\rowcolor{cai_light}
\multicolumn{6}{l}{\textbf{\textcolor{cai_primary}{Hypershell X --- Powered Exoskeleton}}} \\
\midrule

HS-01 & Hypershell & No BLE authentication: any BLE client can connect and send commands via Nordic UART Service without pairing & CWE-306 & 9.8 & Critical \\

HS-02 & Hypershell & IDOR: device IDs derived from reversed BLE MAC address bytes, enabling passive identification from BLE advertisements & CWE-639 & 9.9 & Critical \\

HS-03 & Hypershell & IDOR: API endpoints expose owner email, usage history, battery data, and device binding status for arbitrary device IDs & CWE-639 & 9.9 & Critical \\

HS-04 & Hypershell & BLE pairing secret is the \texttt{motorVersion} field from server API, accessible to any authenticated user for any device (code analysis) & CWE-798 & 9.5 & Critical \\

HS-05 & Hypershell & Unsigned OTA firmware updates; only CRC16 verification; firmware binaries publicly accessible at predictable URLs & CWE-494 & 9.8 & Critical \\

HS-06 & Hypershell & No per-command authentication: once BLE connected, all 177 enumerated commands (including motor control, OTA, reset) are accepted & CWE-862 & 9.9 & Critical \\

HS-07 & Hypershell & Plaintext root MySQL credentials for both Chinese and international servers identified in application artifacts & CWE-798 & 9.9 & Critical \\

HS-08 & Hypershell & SMTP/IMAP credentials (\texttt{it@hypershell.tech}) identified in heap dumps; confirmed access to support mailbox ($\sim$3{,}300 emails) & CWE-798 & 9.7 & Critical \\

HS-09 & Hypershell & Feishu API token confirmed functional; provides access to support database with 64+ customer tickets and user device mappings & CWE-798 & 9.6 & Critical \\

HS-10 & Hypershell & Debug mode accessible in production via \texttt{SET\_DEBUG\_KEYID} command (0xAE); exposes internal protocol details & CWE-489 & 8.5 & High \\

HS-11$^\dagger$ & Hypershell & Motor parameter integer overflow: \texttt{int8ToUint8} conversion causes values $-128$ to $-1$ to become $128$--$255$ (static analysis, no working PoC) & CWE-190 & 8.0 & High \\

HS-12$^\dagger$ & Hypershell & Unvalidated array allocation from BLE packet data: crafted length field may cause heap exhaustion (static analysis, no working PoC) & CWE-120 & 8.2 & High \\

\midrule
\rowcolor{cai_light}
\multicolumn{6}{l}{\textbf{\textcolor{cai_primary}{HOBOT S7 Pro --- Window Cleaning Robot}}} \\
\midrule

HB-01 & HOBOT & No BLE authentication: device accepts connections without pairing, bonding, or passkey; all GATT services immediately accessible & CWE-306 & 9.1 & Critical \\

HB-02 & HOBOT & Unauthenticated BLE command injection: 16 commands validated via PoC including motor control, movement, and factory reset & CWE-862 & 9.1 & Critical \\

HB-03 & HOBOT & Unauthenticated Silicon Labs OTA service accepts arbitrary firmware data writes; no cryptographic signature verification & CWE-494 & 9.8 & Critical \\

HB-04 & HOBOT & Firmware download over plaintext HTTP from \texttt{hobot.com.tw}; enables potential MITM firmware replacement & CWE-319 & 7.4 & High \\

HB-05 & HOBOT & Hardcoded Gizwits cloud credentials (\texttt{app\_id}, \texttt{app\_secret}, \texttt{ProductKey}, \texttt{ProductSecret}) in APK enabling anonymous API access & CWE-798 & 7.5 & High \\

HB-06 & HOBOT & XOR-only integrity check on BLE commands; single byte XOR provides no authentication against adversarial modification & CWE-328 & 7.5 & High \\

HB-07 & HOBOT & No replay protection: identical commands accepted and processed repeatedly in testing; no nonce, counter, or timestamp & CWE-294 & 7.5 & High \\

HB-08 & HOBOT & Device data extraction via unauthenticated BLE: serial number, factory calibration, sensor thresholds, motor logs (135KB+) & CWE-200 & 7.5 & High \\

HB-09 & HOBOT & No GDPR data subject rights implementation: 18 endpoint patterns tested for Art.\ 15/17/20 compliance, all returned HTTP 404 & CWE-359 & N/A & High \\

HB-10 & HOBOT & Excessive data collection via Gizwits cloud: real-time position tracking every 0.7s, floor plans, occupancy indicator, schedules & CWE-359 & N/A & High \\

HB-11 & HOBOT & BLE connection denial of service: single-connection limit allows attacker to monopolize BLE link, denying owner access & CWE-400 & 6.5 & Medium \\

HB-12 & HOBOT & Device information disclosure via GATT: manufacturer, hardware revision, firmware version, system ID exposed without authentication & CWE-200 & 5.3 & Medium \\

HB-13 & HOBOT & SWD debug port likely unlocked on EFR32BG22 SoC (inferred from Silicon Labs default configuration; not empirically verified) & CWE-1191 & 6.8 & Medium \\

HB-14 & HOBOT & Cloud account security deficiencies: unlimited anonymous account creation (20 in 10s observed), 180-day token lifetime, user enumeration & CWE-307 & 6.5 & Medium \\

HB-15$^\ddagger$ & HOBOT & Cross-product data leakage on Gizwits platform: HOBOT credentials retrieved schema of unrelated vendor's device (observed once, not reproducible) & CWE-284 & 5.3 & Medium \\

HB-16 & HOBOT & No application code obfuscation: full BLE protocol, command map, and cloud credentials extractable via jadx decompilation & CWE-656 & 3.7 & Low \\

HB-17 & HOBOT & Outdated BLE module firmware (V0.3.1, April 2022); Silicon Labs Gecko SDK has received multiple security patches since & CWE-1104 & 3.7 & Low \\

\end{longtable}
\end{small}

\arrayrulecolor{black}

\end{document}